\documentstyle[prd,aps,epsfig]{revtex}
\begin{document}
\begin{center}
\boldmath
{\Large\bf Form factors for rare decays $B\to(\pi,\rho,K,K^*)l^+l^-$ in quark
model}
\\
\vspace{.5cm}
{\large D.~Melikhov and N.~Nikitin}
\\
\vspace{.5cm}
{\normalsize {\it Nuclear Physics Institute, Moscow State University, Moscow,
119899, Russia} \\ 
Electronic address: melikhov@monet.npi.msu.su, nik679@sgi.npi.msu.su}
\end{center}
\unboldmath

\begin{abstract}
Hadronic matrix elements of the vector, axial--vector and tensor currents 
relevant for exclusive rare decays $B\to(\pi,\rho,K,K^*)\,l^+l^-$ induced 
by the quark transition $b\to d,s$ are
analysed using dispersion formulation of the light--cone constituent quark
model. The form factors in the decay region are expressed as relativistic 
double spectral representations through the light--cone wave functions
of the initial and final mesons. 
These form factors are shown to have the correct scaling properties in the
limits of  
heavy--to--heavy and heavy--to--light quark transitions in
accordance with QCD. The dependence on the quark--model parameters is studied. 
\end{abstract}

\vspace{.5cm}

Rare decays of g$\bar B$ mesons induced by the flavour--changing neutral current transition 
$b\to s,d$ provide an important probe of the Standard Model (SM) and its extentions. 
These decays are forbidden at the tree level  
and occur in the lowest order through one--loop diagrams and thus open a possibility 
to probe at comparatively low energies the structure of the theory at high mass scales 
which shows through virtual particles in the loops. 
The amplitudes of exclusive $B$ meson decay modes involve the nonperturbative 
long--distance contributions which should be known with high accuracy to reliably separate 
the short--distance effects. 
The main long--distance contribution in the decay of $B$--mesons is given 
by the transition amplitudes of the bilinear quark currents. 
A theoretical study of these amplitudes encounters the problem of describing the hadron 
structure and requires a nonperturbative consideration. This gives the main uncertainty to the 
theoretical predictions for hadron transition amplitudes. 

For the calculation of the amplitudes of exclusive rare meson decays 
one needs matrix elements of the vector, axial--vector, and tensor currents which have the following 
stucture \cite{iw1}
\begin{eqnarray}
\label{amplitudes}
<P(M_2,p_2)|V_\mu(0)|P(M_1,p_1)>&=&f_+(q^2)P_{\mu}+f_-(q^2)q_{\mu},  \nonumber \\
<V(M_2,p_2,\epsilon)|V_\mu(0)|P(M_1,p_1)>&=&2g(q^2)\epsilon_{\mu\nu\alpha\beta}
\epsilon^{*\nu}\,p_1^{\alpha}\,p_2^{\beta}, \nonumber \\
<V(M_2,p_2,\epsilon)|A_\mu(0)|P(M_1,p_1)>&=&
i\epsilon^{*\alpha}\,[\,f(q^2)g_{\mu\alpha}+a_+(q^2)p_{1\alpha}P_{\mu}+
               a_-(q^2)p_{1\alpha}q_{\mu}\,],   \nonumber \\
<P(M_2,p_2)|T_{\mu\nu}(0)|P(M_1,p_1)>&=&-2i\,s(q^2)\,(p_{1\mu}p_{2\nu}-p_{1\nu}p_{2\mu}), \nonumber \\
<V(M_2,p_2,\epsilon)|T_{\mu\nu}(0)|P(M_1,p_1)>&=&i\epsilon^{*\alpha}\,
               [\,g_{+}(q^2)\epsilon_{\mu\nu\alpha\beta}P^{\beta}+
                g_{-}(q^2)\epsilon_{\mu\nu\alpha\beta}q^{\beta}+
                h(q^2)p_{1\alpha}\epsilon_{\mu\nu\beta\gamma}p_1^{\beta}p_2^{\gamma}\,], \nonumber \\
<P(M_2,p_2)|T^5_{\mu\nu}(0)|P(M_1,p_1)>&=&s(q^2)\,\epsilon_{\mu\nu\alpha\beta}P^\alpha q^\beta ,
\nonumber \\
<V(M_2,p_2,\epsilon)|T^5_{\mu\nu}(0)|P(M_1,p_1)>&=&
g_+(q^2)(\epsilon_\nu^*P_\mu-\epsilon_\mu^*P_\nu) 
+g_-(q^2)(\epsilon_\nu^*q_\mu-\epsilon_\mu^*q_\nu) \nonumber \\
&&+h(q^2)(p_1\epsilon^*)(p_{1\nu}p_{2\mu}-p_{1\mu}p_{2\nu})
\end{eqnarray}
where the quark currents have the following form
$$
V_{\mu}\;=\;{\bar q}\gamma_{\mu}b, \qquad 
A_{\mu}\;=\;{\bar q}\gamma_{\mu}\gamma_{5}b,\qquad 
T_{\mu\nu}\;=\;{\bar q}\sigma_{\mu\nu}b,\qquad
T^5_{\mu\nu}\;=\;{\bar q}\sigma_{\mu\nu}\gamma_5b,
$$
$q=p_{1}-p_{2}$, $P=p_{1}+p_{2}$
\footnote{We use the following notations: 
$\gamma^{5}=i\gamma^{0}\gamma^{1}\gamma^{2}\gamma^{3}$, 
$\sigma_{\mu \nu}={\frac{i}{2}}[\gamma_{\mu},\gamma_{\nu}]$,
$\epsilon^{0123}=-1$,  
$\gamma_{5}\sigma_{\mu\nu}=-{\frac{i}{2}}\epsilon_{\mu\nu\alpha\beta}\sigma^{\alpha\beta}$, and 
$Sp(\gamma^{5}\gamma^{\mu}\gamma^{\nu}\gamma^{\alpha}\gamma^{\beta})=
4i\epsilon^{\mu\nu\alpha\beta}$.}.

The relativistic--invariant form factors which contain the dynamical 
information on the process should be calculated within a nonperturbative approach 
for any particular initial and final mesons. 
However, in some cases simplifications occur which reduce the number of the independent 
form factors. \\
{\bf 1. Meson transition induced by the transition between heavy quarks.} \\
Due to the heavy quark symmetry \cite{iw2} all the form factors which are in general 
functions of $q^2$ and the masses, 
depend on the one dimensionless variable $\omega=p_1p_2/M_1M_2$ and in the leading $1/m_Q$ order 
can be expressed through the single universal form factor, the Isgur--Wise function as 
follows 
\cite{iw2}:
\begin{eqnarray}
\label{hh}
f_+(\omega)&=&-g_+(\omega)=\frac{M_1+M_2}{2\sqrt{M_1M_2}}\xi(\omega)\nonumber \\
f_-(\omega)&=&-g_-(\omega)=-\frac{M_1-M_2}{2\sqrt{M_1M_2}}\xi(\omega)\nonumber \\
s(\omega)&=&g(\omega)=a_-(\omega)=-a_+(\omega)=\frac{1}{2\sqrt{M_1M_2}}\xi(\omega)\nonumber \\
f(\omega)&=&\sqrt{M_1M_2}(1+\omega)\xi(\omega)\nonumber\\
h(\omega)&=&0.
\end{eqnarray}
In addition, the normalization of the Isgur--Wise function at zero recoil is known, namely $\xi(1)=1$. \\
{\bf 2. Meson transition caused by the heavy--to--light quark transition.}\\
In this case the form factors of the tensor current in the region $\omega=O(1)$ 
can be expressed through the form factors of the vector and axial--vector 
currents in the leading $1/m_Q$ 
order [$M_1=m_Q+O(1)$] as follows \cite{iw1}: 
\begin{eqnarray}
\label{hla}
s(q^2)&=&\frac{1}{2M_1}[f_+(q^2)-f_-(q^2)],  \\
\label{hlb}
h(q^2)&=&\frac{1}{M_1}[2g(q^2)+a_{+}(q^2)-a_{-}(q^2)], \\
\label{hlc}
g_{-}(q^2)-g_{+}(q^2)&=&2M_1\,g(q^2), \\
\label{hld}
g_{-}(q^2)+g_{+}(q^2)&=&\frac{1}{M_1}[-f(q^2)+2p_1p_2\cdot g(q^2)]. 
\end{eqnarray}

All the rest unknown form factors as well as the Isgur--Wise function in the case of the
heavy--to--heavy transition should be determined within a dynamical approach.

An important step forward in the understanding of the heavy meson decay form factors has been
recently done by B.~Stech \cite{stech} who noticed that new relations between the form 
factors of meson transition can be derived if use is made of the constituent quark
picture. 
For the case of the heavy--to--heavy quark transition these relations reproduce the 
Isgur--Wise relations (\ref{hh}), and for the case of the heavy--to--light quark 
transition in addition to the relations (\ref{hla}--\ref{hld}) 
new constraints on the ratios of the form factors occur. 
They are based on the assumption that the meson wave function in terms of its 
quark constituents is strongly peaked in the momentum space with the width of order 
$\Lambda\simeq 0.5\; GeV$. In this case a natural small parameter $\Lambda/m_Q$ appears 
in the picture and one can derive the
leading--order expressions for the form factors of interest which turn out to be
independent on subtle details of the meson structure. 
Although these relations considerably simplify the analysis of heavy--to--light transitions,
the corrections to the leading behavior are not under control (their 
calculation requires more details of the meson structure and dynamics of the decay process) 
and it seems necessary to estimate the form factors in a particular dynamical model. 

Usually the form factors are investigated within such nonperturbative methods as 
various versions of the quark model \cite{wsb,isgw,jaus,jw,beyer,faustov,demchuk,cheng}, 
QCD sum rules \cite{sr1,sr2,sr3}
(more references can be found in \cite{sr4})
and lattice QCD \cite{lat1,lat2}. 

As the kinematically accessible $q^2$--interval in the meson decay induced by the 
heavy--to--light transition is $O(m_Q^2)$, the consideration is mandatory to be 
relativistic. 

The relativistic light--cone quark model (LCQM) \cite{jaus} is an adequate framework for 
considering such decays. The difficulty with the direct application of the model at timelike 
momentum transfer $q$ 
lies in the contribution of pair--creation subprocesses which at $q^2>0$ cannot be killed 
by an appropriate choice of the reference frame (see the discussion and references in \cite{m2}).
In \cite{jaus,jw} the form factors at $q^2>0$ were obtained by numerical extrapolation from the 
region $q^2<0$. As the relevant $q^2$--interval is large, the accuracy of such a procedure is not
high. The dispersion formulation of the LCQM proposed in \cite{m1} overcomes this difficulty as it 
allows the analytical continuation to the timelike $q$. We apply this approach
to calculating the transition form factors (\ref{amplitudes}) for the decays 
$B\to f$ where $f=\pi,\rho,K,K^*$. 
 
The transition of the initial meson  $Q(m_2)\bar Q(m_3)$ with the mass $M_1$ 
to the final meson $Q(m_2)\bar Q(m_3)$ with the mass $M_2$ 
induced by the quark transition $m_2\to m_1$ is described by the diagram of Fig.1. 
The constituent quark structure of the initial and final mesons are described by the vertices $\Gamma_1$
and $\Gamma_2$, respectively. The initial pseudoscalar meson the vertex has the spinorial structure
$\Gamma_1=i\gamma_5G_1/\sqrt{N_c}$; the final meson vertex has the structure 
$\Gamma_2=i\gamma_5G_2/\sqrt{N_c}$ for a pseudoscalar state and the structure 
$\Gamma_{2\mu}=[A\gamma_\mu+B(k_1-k_3)_\mu]\,G_2/\sqrt{N_c}$ for the vector state. The values 
$A=-1$, $B=1/(\sqrt{s_2}+m_1+m_3)$ correspond to an $S$--wave vector meson, and 
$A=1/\sqrt{2}$, $B=[2\sqrt{s_2}+m_1+m_3]/[\sqrt{2}(s_2+(m_1+m_3)^2)]$ 
correspond to a $D$--wave vector particle. 

We start with the region $q^2<0$ and make use of the fact that the form factors 
of the LCQM \cite{jaus,jw} can be represented 
as double spectral representations over the invariant masses of the initial and final
$q\bar q$ pairs as follows \cite{m1}
\begin{equation}
\label{ff}
f_i(q^2)=f_{21}(q^2)
\int\limits^\infty_{(m_1+m_3)^2}\frac{ds_2G_{2}(s_2)}{\pi(s_2-M_2^2)}
\int\limits^{s_1^+(s_2,q^2)}_{s_1^-(s_2,q^2)}\frac{ds_1G_{1}(s_1)}{\pi(s_1-M_1^2)}
\frac{\tilde f_i(s_1,s_2,q^2)}{16\lambda^{1/2}(s_1,s_2,q^2)},
\end{equation}
where 
$$
s_1^\pm(s_2,q^2)=
\frac{s_2(m_1^2+m_2^2-q^2)+q^2(m_1^2+m_3^2)-(m_1^2-m_2^2)(m_1^2-m_3^2)}{2m_1^2}
\pm\frac{\lambda^{1/2}(s_2,m_3^2,m_1^2)\lambda^{1/2}(q^2,m_1^2,m_2^2)}{2m_1^2}
$$
and 
$
\lambda(s_1,s_2,s_3)=(s_1+s_2-s_3)^2-4s_1s_2
$
is the triangle function. Here
$f_{21}(q^2)$ is the form factor of the constituent quark transition $m_2\to m_1$. 
In what follows we set $f_{21}(q^2)=1$. 
A similar representation of the form factors has been obtained in \cite{akms} 
taking into account the contribution of two--particle singularities in the 
Feynman graphs. The double spectral densities $\tilde f_i(s_1,s_2,q^2)$ of the form factors 
are calculated from the Feynman graph of Fig.1 with all intermediate particles on the mass shell
and pointlike vertices corresponding to the initial and final mesons and have the 
following form: 
\begin{eqnarray}
\label{fplus}
&&\tilde f_++\tilde f_-=4\;[m_1m_2\alpha_1-m_2m_3\alpha_1+m_1m_3(1-\alpha_1)-m_3^2(1-\alpha_1)
+\alpha_2s_2],
\\
\label{fminus}
&&\tilde f_+-\tilde f_-=4\;[m_1m_2\alpha_2-m_1m_3\alpha_2+m_2m_3(1-\alpha_2)-m_3^2(1-\alpha_2)
+\alpha_1s_1],
\\
\label{g}
&&\tilde g=-2A\,[m_1\alpha_{2}+m_2\alpha_{1}+m_3(1-\alpha_{1}-\alpha_{2})]-4B\beta,
\\
\label{a1}
&&\tilde a_++\tilde a_-=-4A\,[2m_2\alpha_{11}+2m_3(\alpha_1-\alpha_{11})]+4B[C_1\alpha_1+C_3\alpha_{11}],
\\
\label{a2}
&&\tilde a_+-\tilde a_-=
-4A\,[-m_1\alpha_{2}-m_2(\alpha_{1}-2\alpha_{12})-m_3(1-\alpha_1-\alpha_2+2\alpha_{12})]
+4B\,[C_2\alpha_1+C_3\alpha_{12}],
\\
\label{f}
&&\tilde f=\frac{M_2}{\sqrt{s_2}}\tilde f_{D}+
\left({\frac{s_1-s_2-s_3}{2\sqrt{s_2}}-\frac{M_1^2-M_2^2-s_3}{2M_2}}\right)M_2\tilde a_+, 
\\
\label{s}
&&\tilde s=2\,[m_1\alpha_{2}+m_2\alpha_{1}+m_3(1-\alpha_{1}-\alpha_{2})],
\\
\label{gplus}
&&\tilde g_{+}+\tilde g_{-}=4A\,[m_3(m_1-m_3)+\alpha_1(m_1-m_3)(m_2-m_3)+\alpha_2s_2+2\beta]
+8B(m_1+m_3)\beta,
\\
\label{gminus}
&&\tilde g_{+}-\tilde g_{-}=4A\,[m_3(m_2-m_3)+\alpha_2(m_1-m_3)(m_2-m_3)+\alpha_1s_1]
+8B(m_2-m_3)\beta,
\\
\label{h}
&&\tilde h=-8A\alpha_{12}
-8B\,[-m_3\alpha_{1}+(m_3-m_2)\alpha_{11}+(m_3+m_1)\alpha_{12}],
\end{eqnarray}
where 
\begin{eqnarray}
\label{fdisp}
\tilde f_D&=&-4A[m_1m_2m_3+\frac{m_2}2(s_2-m_1^2-m_3^2)
+\frac{m_1}2(s_1-m_2^2-m_3^2)-\frac{m_3}2(s_3-m_1^2-m_2^2)\\
&&+2\beta(m_2-m_3)]+4B\,C_3\beta,\nonumber
\end{eqnarray}

\begin{eqnarray}
\label{alpha1}
&\alpha_1=\left[(s_1+s_2-s_3)(s_2-m_1^2+m_3^2)-2s_2(s_1-m_2^2+m_3^2)\right]/{\lambda(s_1,s_2,s_3)},
\\
\label{alpha2}
&\alpha_2=
\left[(s_1+s_2-s_3)(s_1-m_2^2+m_3^2)-2s_1(s_2-m_1^2+m_3^2)\right]/{\lambda(s_1,s_2,s_3)},
\\
\label{beta}
&\beta=\frac14\left[2m_3^2-\alpha_1(s_1-m_2^2+m_3^2)-\alpha_2(s_2-m_1^2+m_3^2)\right],
\\
\label{alpha11}
&\alpha_{11}=\alpha_1^2+4\beta {s_2}/{\lambda(s_1,s_2,s_3)}, \quad
\alpha_{12}=\alpha_1\alpha_2-2\beta(s_1+s_2-s_3)/\lambda(s_1,s_2,s_3),
\\
\label{cc}
&C_1=s_2-(m_1+m_3)^2, \quad C_2=s_1-(m_2-m_3)^2, \quad C_3=s_3-(m_1+m_2)^2-C_1-C_2.
\end{eqnarray}

Let us underline that the representation (\ref{ff}) with the spectral densities
(\ref{fplus}--\ref{h}) are just 
the dispersion form of the corresponding light--cone expressions from \cite{jaus,jw}. 
It is important that double spectral representations without subtractions are 
valid for all the form factors except $f$ which requires subtractions
\footnote{The quantity $\tilde f_D$ is calculated directly from the double--cut Feynman graph of Fig.1
as explained in \cite{m1}. The spectral representation is constructed from $\tilde f_D$ by means of a
subtraction procedure. In the LCQM the particular form of such a representation for the form factor $f$
depends on the choice of the current component used for its determination.}.
 
Both for a pseudoscalar and vector meson ($S$ and $D$--wave) with the mass $M$ built up of the 
constituent quarks $m_q$ and $m_{\bar q}$, the function $G$ is normalized as 
follows \cite{m1}
\begin{equation}
\label{norma}
\int\frac{G^2(s)ds}{\pi(s-M^2)^2}\frac{\lambda^{1/2}(s,m_q^2,m_{\bar q}^2)}{8\pi s}
(s-(m_q-m_{\bar q})^2)=1. 
\end{equation}

As the analytical continuation of the from factors (\ref{ff}) to the timelike region is performed, 
in addition to the normal contribution which is just the expression (\ref{ff}) taken
at $q^2>0$ an additional anomalous contribution emerges. The corresponding expression is given in
\cite{m1}. The normal contribution dominates the form factor at small timelike and vanishes as 
$q^2=(m_2-m_1)^2$ while the anomalous contribution is negligible at small $q^2$ and steeply rises
as $q^2\to(m_2-m_1)^2$. 

Let us study the form factors given by (\ref{ff}) with the spectral densities
(\ref{fplus}--\ref{h}) in the limit of a heavy initial meson ($m_2\to\infty$) for heavy and
light final states. As the underlying process for the meson decay is the quark--quark transition it is convenient to
introduce a new variable $\omega_q=(m_1^2+m_2^2-q^2)/2m_1m_2$ which turns out to be more appropriate
for the description of the decay in the quark model than the kinematical variable $\omega$. 
\\
{\bf 1. Heavy--to--heavy quark transition}. \\
In this case $m_{1,2}\to\infty$, $m_1/m_2=O(1)$. 
For the heavy--to--heavy quark transition $\omega=\omega_q+O(1/m_Q)$ where $m_Q$ is either $m_1$ or $m_2$.

The soft vertex functions $G$ are strongly peaked near the $q\bar q$ threshold. 
We introduce a new variable $z$ such that $s=(m_Q+m_{\bar q}+z)^2$ and assume that 
$G(z)$ is peaked in the region $z\le \Lambda$. 
If we consider the heavy meson case ($m_Q>>m_{\bar q}\simeq \Lambda \simeq \epsilon$; 
$M=m_Q+m_{\bar q}-\epsilon)$ then the normalization
condition (\ref{norma}) takes the from 
\begin{equation}
\label{normaQ}
\int\varphi^2(z)(2m_{\bar q}+z)\sqrt{z(2m_{\bar q}+z)}=1+O(1/m_Q)
\end{equation}
with 
$$
\varphi(z)=\frac{G(z)}{2\pi \sqrt{m_Q}(z+\epsilon)}.
$$

The anomalous contribution to the transition form factors is suppressed by additional powers of
$1/m_Q$ everywhere except the point $\omega_q=1$. 
Then at $\omega_q\ne 1$ one recovers the relations (\ref{hh}) with the Isgur--Wise function  
\begin{eqnarray}
&\label{xi}
\xi(\omega_q)=\int\limits_0^\Lambda 
dz_2 \varphi_2(z_2)\sqrt{z_2(z_2+2m_3)}\int\limits_{-1}^{1}\frac{d\eta}2
\varphi_1(z_1)\left({m_3+\frac{2m_3+z_1+z_2}{1+\omega_q}}\right); \\
&z_1=z_2\omega_q+m_3(\omega_q-1)+\eta\sqrt{z_2(z_2+2m_3)}\sqrt{\omega_q^2-1}+O(1/m_Q). \nonumber
\end{eqnarray}
Since $\xi$ is a continuous function, the representation (\ref{xi}) is valid also at $\omega_q=1$. 
The heavy mesons have the same wave functions ($\varphi_1=\varphi_2$) 
and hence $\xi(1)=1$ as follows from (\ref{norma}). 

Let us point out that the particular form of the subtraction term in the form factor $f$ 
is not important for its large--$m_Q$ behaviour as this term has the next--to--leading 
$1/m_Q$ order. However, this is not the case when the final quark is light. \\
{\bf 2. Heavy--to--light quark transition}. \\
In this case $m_2\to\infty$ with $m_1,m_3$ and $\Lambda$ kept finite. 
One can find the spectral densities to obey in the leading $1/m_2$ order the 
following relations (both for $\omega=O(1)$ and $\omega=O(m_2)$):
\begin{eqnarray}
\label{hl1}
\tilde s&=&\frac{1}{2m_2}(\tilde f_+-\tilde f_-), \\
\label{hl2}
\tilde h&=&
\frac{1}{m_2}(2\tilde g+\tilde a_{+}-\tilde a_{-}), \\
\label{hl3}
\tilde g_{-}-\tilde g_{+}&=&2m_2\tilde g,  \\
\label{hl4}
\tilde g_{-}+\tilde g_{+}&=&\frac{1}{m_2}[-\tilde f_D+(s_1+s_2-q^2)\tilde g].
\end{eqnarray}

These relations yield the fulfillment of the expressions (\ref{hla}--\ref{hlc}) 
for the form factors 
given by the dispersion representations without subtractions. However, the 
form factor $f$ with the spectral density (\ref{f}) 
does not satisfy the relation (\ref{hld}). 
To obtain the form factor $f$ which has the correct large--$m_2$ behaviour, we 
redefine the subtraction procedure as follows
\footnote{Numerically the results obtained with the form factors given 
by (\ref{f}) and (\ref{fhl}) are different by at most 
10\% for the $B\to K^*$ and $B\to\rho$ transitions.}:
\begin{equation}
\label{fhl}
\tilde f(s_1,s_2,q^2)=\tilde f_D(s_1,s_2,q^2)+(M_1^2-s_1+M_2^2-s_2)\tilde g(s_1,s_2,q^2).
\end{equation}

Let us turn to the scaling properties of the form factors in the limit $m_2\to\infty$.
In the region $\omega_q=O(1)$, i.e. $q^2=O(m_2^2)$, one finds for the normal contribution
(the anomalous term has a similar behaviour)

\begin{equation}
\label{scaling}
f_i(\omega_q)
=
\frac{1}{\sqrt{m_2}}
\int
\frac{ds_2G_2(s_2)\lambda^{1/2}(s_2,m_1^2,m_3^2)}
{16\,m_1\pi(s_2-M_2^2)}
\int\limits_{-1}^{1}
\frac
{d\eta \sqrt{\omega_q^2-1}\;\varphi(z_1)\tilde f_i(z_1,z_2,m_1,m_2,m_3,\omega_q)}
{\sqrt{(m_1(\omega_q+1)+z_1+z_2+2m_3)(m_1(\omega_q-1)+z_1-z_2)}}
\end{equation}
where
$$
z_1=z_2\omega_q\left({1+\frac{2m_3+z_2}{m_1}}\right)+m_3(\omega_q-1)+\eta\sqrt{\omega_q^2-1}
\frac{\lambda^{1/2}(s_2,m_1^2,m_3^2)}{2m_1}+O(1/m_2).
$$
The spectral densities scale at large $m_2$ as  
\begin{equation}
\tilde f_i=m_2^{n_i}\rho_i(\omega_q,m_1,m_3,z_1,z_2).
\end{equation}
Namely, 
\begin{eqnarray}
&\tilde f_++\tilde f_-=O(1),\quad 
\tilde f_+-\tilde f_-=O(m_2),\quad 
\tilde g=O(1),\quad 
\tilde a_++\tilde a_-=O(1/m_2),\quad 
\tilde a_+-\tilde a_-=O(1),
\nonumber
\\
&\tilde f=O(m_2),\quad 
\tilde g_++\tilde g_-=O(1),\quad 
\tilde g_+-\tilde g_-=O(m_2),\quad 
\tilde h=O(1/m_2),\quad 
\tilde s=O(1).
\nonumber
\end{eqnarray}
Hence, the form factors have the scaling behaviour of the form
\begin{equation}
f_i=m_2^{n_i-1/2}r_i(\omega_q;m_1,m_3,\epsilon_2,G_2;\varphi)
\end{equation}
where $\varphi$ is a universal function for all heavy mesons. 
The variables $\omega$ and $\omega_q$ are connected with each other as follows
\begin{equation}
\omega_q=\omega\left({1+\frac{m_3-\epsilon_2}{m_1}}\right)+\frac{m_3-\epsilon_1}{m_1}+O(1/m_2).
\end{equation}
If the quark masses are chosen such that in a heavy meson $Q\bar q$ with the mass $M$ 
the binding energy has the form 
$\epsilon(m_Q)=\epsilon+O(1/m_Q)$, then the ratio
\begin{equation}
f_i(\omega)/M^{n_i+1/2}
\end{equation}
is universal for the transition of any heavy meson into a fixed  light state. 
This behaviour reproduces the results of \cite{iw1} up to the
logarithmic corrections which arise from the anomalous scaling of the quark currents in QCD. 

Summing up, we have the representations for the form factors which reproduce the correct 
behaviour in the two important cases of a heavy meson transition into heavy and light final states. 
We are in a position now to apply these representations to calculating the form factors for 
the $B\to\pi,\rho,K,K^*$ decays.

We parametrise the quark--meson vertices as 
\begin{equation}
\label{5vertex}
G(s)=\frac{\pi}{\sqrt{2}}\frac{\sqrt{s^2-(m_1^2-m_2^2)^2}}
{\sqrt{s-(m_1-m_2)^2}}\frac{s-M^2}{s^{3/4}}w(k),\qquad
k=\frac{\lambda^{1/2}(s,m_1^2,m_2^2)}{2\sqrt{s}}
\end{equation}
and following \cite{jaus} assume the exponential dependence of the ground--state $S$--wave 
light--cone radial wave function $w(k)$: 
\begin{equation}
\label{exp}
w(k)=\exp(-k^2/2\beta^2)
\end{equation}

We ran calculations with several sets of the quark model parameters 
used for the description of meson mass spectrum and elastic form factors 
(Table \ref{table:parameters}).
Table \ref{table:ffs} gives the parameters of the fits to the calculated form
factors.  
All the form factors except $f$ are fitted by the function
$$
f_{i}(q^2)=f_{i}(0)/(1-q^2/M_i^2)^{n_i};
$$ 
for $f$ another fit is used: 
$$
f(q^2)=f(0)/[1-q^2/M_1^2+(q^2/M_2^2)^2].
$$ 
The radiative transition to the vector state is given by the matrix element 
of the magnetic penguin operator \cite{heff}  
\begin{eqnarray}
&<V(M_2,p_2,\epsilon)|\bar q\sigma_{\mu\nu}q^\nu(1+\gamma_5)b|P(M_1,p_1)>=
i\epsilon^{*\alpha}\epsilon_{\mu\nu\alpha\beta}q^\nu P^{\beta}g_{+}(q^2)   
\nonumber \\
\label{penguin}
&
+(P_\mu\,\epsilon^*q-\epsilon^*_\mu Pq)\left({g_++\frac{q^2}{Pq}g_-}\right)
+(\epsilon^*q)(q_\mu\,Pq-P_\mu q^2)\left({-\frac12 h+\frac{1}{Pq}g_-}\right).
\end{eqnarray}
Only $g_+(0)$ contributes to the amplitude of the transition to the real photon. 
Table \ref{table:bvg} compares our results on $g_+(0)$ 
with recent predictions of other models. 

The Isgur--Wise relations for the form factors of the tensor and vector currents 
(\ref{hla}--\ref{hld}) are fulfilled with an accuracy of $5\div10\%$ for all the considered 
transitions in the whole kinematical interval. We see in an explicit calculation that the
dominance of the soft contribution actually extends the fulfillment of these relations 
originally derived for the region $\omega=O(1)$ to the whole kinematical region. 

Our results are in a good agreement with the model \cite{stech}; on the other hand
one finds considerable difference from the QCD sum rule calculations \cite{sr3} 
in some of the $B\to K,K^*$ form factors (see Table \ref{table:ffs0}). 

One can observe rather strong dependence of the form factors on the parameters of the quark model. 
Several well--measured points of the decay spectrum could be helpful in choosing the values relevant for
the decay processes \cite{stech}. 

In addition to the long--distance effects in the amplitudes of the 
bilinear quark currents desribed by the calculated form factors, 
the amplitude of the rare semileptonic decays contains also 
other long--distance effects induced by the four--quark operators in the 
effective Hamiltonian.  

One of these effects is the contribution to the effective Hamiltonian from
the $c\bar c$ resonances in the $q^2$--channel which appear in the physical region of the $B\to K,K^*$
decay. This contribution has been described by introducing 
the effective Wilson coefficient ${\cal C}_7^{eff}$ \cite{ali} which includes the effects of 
renormalization 
of the penguin operator and both the short and the long distance effects of the four--quark
operators. Within the constituent quark picture this long--distance contribution in the 
$q^2$--channel should be rather attributed to the constituent quark form factor. 
The long--distance effects of the same type take place also in semileptonic decays 
where meson states with relevant quantum numbers give rise to a nontrivial structure of the 
constituent quark transition form factor \cite{m1}. A specific feature of rare decays is the appearance 
of the resonances directly in the kinematical region of meson decay, whereas in 
semileptonic decays the resonances are above this region. Thus in rare decays the contribution
of these resonances can be more important. 

The second effect is the weak annihilation \cite{long1} which has been estimated to give a 
few percent correction in some of the decay modes,
whereas in other decays it has been claimed to be the main effect. 
Both of these long--distance contributions should be taken into account for a realistic calculation 
of the leptonic spectra and the decay rates. We would like to address this issue elsewhere. 

We are grateful to L.~Kondratyuk, I.~Narodetskii, S.~Simula, and K.~Ter--Martirosyan 
for discussions.
The work was supported by the Russian Foundation for Basic Researh under 
grant 96--02--18121a.

\begin{figure}[1]
\begin{center}  
\mbox{\epsfig{file=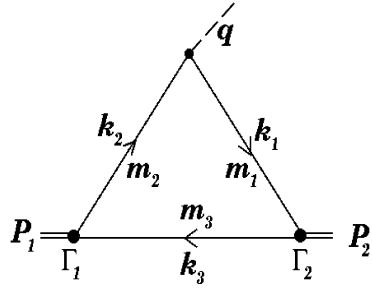,height=4.cm}  }
\end{center}
\caption{One-loop graph for a meson decay.\label{fig:1}}
\end{figure}

\begin{table}[1]
\caption{\label{table:parameters}
Parameters of the quark model.}
\centering
\begin{tabular}{|c||c|c|c|c|c|c|c|c|c|c|}
Ref. & $m_u$ & $m_s$ &$m_b$ & 
$\beta_{\pi}$ & $\beta_{\rho}$ & $\beta_{K}$ & $\beta_{K^{*}}$ & 
$\beta_{\phi}$ & $\beta_{B_u}$ & $\beta_{B_s}$  \\
\hline
\hline
Set 1 \cite{wsb}     & 0.35  & 0.55 & 4.9  &  0.4   & 0.4   & 0.4   & 0.4   & 0.4   & 0.4   & 0.4  \\
Set 2 \cite{isgw}    & 0.33  & 0.55 & 5.2  &  0.41  & 0.30  & 0.44  & 0.33  & 0.37  & 0.43  & 0.54 \\
Set 3 \cite{jaus}    & 0.29  & 0.42 & 4.9  &  0.291 & 0.291 & 0.358 & 0.358 & $-$   & 0.386 & $-$  \\
Set 4 \cite{schlumpf}& 0.25  & 0.37 & 4.7  &  0.32  & 0.32  & 0.395 & 0.395 & $-$   & 0.55  & $-$  \\
\end{tabular}
\end{table}

\begin{table}[2]
\caption{\label{table:ffs}
Parameters of the fits to the calculated $B\to P,V$ transition form factors.}
\centering
\begin{tabular}{|c|c|c|c|c|c|c|c|c|c|c|}
 Decays      & \multicolumn{3}{c|}{$B \to \pi$} & \multicolumn{7}{c|}{$B \to \rho$}   \\
\hline
\hline
             & $f_{+}(0)$ & $f_{-}(0)$  &$s(0)$  &$g(0)$ &$f(0)$ &$a_{+}(0)$ &$a_{-}(0)$ & $h(0)$  & $g_{+}(0)$  & $g_{-}(0)$      \\  
 Ref.        & $M_{f_{+}}$& $M_{f_{-}}$ &$M_{s}$ &$M_{g}$&$M_{1}$&$M_{a_{+}}$&$M_{a_{-}}$& $M_{h}$ & $M_{g_{+}}$ & $M_{g_{-}}$     \\
             & $n_{f_{+}}$& $n_{f_{-}}$ &$n_{s}$ &$n_{g}$&$M_{2}$&$n_{a_{+}}$&$n_{a_{-}}$& $n_{h}$ & $n_{g_{+}}$ & $n_{g_{-}}$     \\
\hline
\hline
Set 1        & 0.29   & -0.26    & 0.05         & 0.057 & 1.69  & -0.035    &  0.045 & 0.0065 & -0.31 & 0.29   \\
             & 6.48   &  6.34    & 6.47         & 6.47  & 6.20  & 8.48      &  7.22  & 6.39   & 6.48  & 6.38   \\
             & 2.54   &  2.49    & 2.50         & 2.54  & 7.20  & 3.57      &  2.86  & 3.27   & 2.55  & 2.54   \\
\hline
Set 2        & 0.29   & -0.26    & 0.05         & 0.036 & 1.10 & -0.026     &  0.030 & 0.0030 & -0.20 & 0.18   \\
             & 6.71   &  6.55    & 6.68         & 6.55  & 5.59 & 7.29       &  6.88  & 6.43   & 6.57  & 6.50   \\
             & 2.35   &  2.30    & 2.31         & 2.75  & 7.10 & 3.04       &  2.85  & 3.42   & 2.76  & 2.73   \\
\hline
Set 3        & 0.21   & -0.18    & 0.04         & 0.040 & 1.23 & -0.029     &  0.033 & 0.0038 & -0.22 & 0.20   \\
             & 6.14   &  6.05    & 6.13         & 6.07  & 5.27 & 6.93       &  6.47  &  6.00   & 6.09  & 6.02   \\
             & 2.60   &  2.65    & 2.57         & 2.56  & 7.20 & 2.95       &  2.70  &  3.27  & 2.58  & 2.55   \\
\hline
Set 4        & 0.26    & -0.22   & 0.05         & 0.055 & 1.68 & -0.039     &  0.046 & 0.0054 & -0.29 & 0.26   \\
             & 5.64    &  5.55   & 5.61         & 5.66  & 6.10 & 6.47       &  5.93  & 5.64   & 5.67  & 5.62   \\
             & 1.88    &  1.81   & 1.84         & 1.91  & 19.6 & 2.11       &  1.93  & 2.66   & 1.93  & 1.90   \\
\hline
\hline
 Decays      & \multicolumn{3}{c|}{$B \to K$}   & \multicolumn{7}{c|}{$B \to K^{*}$}   \\
\hline
             & $f_{+}(0)$ & $f_{-}(0)$  &$s(0)$  &$g(0)$ &$f(0)$ &$a_{+}(0)$ &$a_{-}(0)$ & $h(0)$  & $g_{+}(0)$  & $g_{-}(0)$      \\  
 Ref.        & $M_{f_{+}}$& $M_{f_{-}}$ &$M_{s}$ &$M_{g}$&$M_{1}$&$M_{a_{+}}$&$M_{a_{-}}$& $M_{h}$ & $M_{g_{+}}$ & $M_{g_{-}}$     \\
             & $n_{f_{+}}$& $n_{f_{-}}$ &$n_{s}$ &$n_{g}$&$M_{2}$&$n_{a_{+}}$&$n_{a_{-}}$& $n_{h}$ & $n_{g_{+}}$ & $n_{g_{-}}$     \\
\hline
\hline
Set 1        & 0.34   &  -0.28      & 0.06       & 0.063 & 2.02 & -0.042 &  0.052 & 0.0062 & -0.35 & 0.31   \\
             & 6.58   &  6.43       & 6.55       & 6.57  & 5.97 & 7.76   &  7.07  & 6.48   & 6.57  & 6.45   \\
             & 2.56   &  2.51       & 2.53       & 2.56  & 7.53 & 3.05   &  2.75  & 3.27   & 2.58   & 2.55   \\
\hline
Set 2        & 0.36   &  -0.30      & 0.06       & 0.048 & 1.61 & -0.036 &  0.041 & 0.0037 & -0.28 & 0.24   \\
             & 6.88   &   6.71      & 6.85       & 6.67  & 5.86 & 7.33   &  6.98  & 6.57   & 6.67  & 6.59   \\
             & 2.32   &   2.27      & 2.28       & 2.61  & 7.66 & 2.85   &  2.72  & 3.28   & 2.62  & 2.58   \\
\hline
Set 3        & 0.30   &  -0.25      & 0.06       & 0.057 & 1.80 & -0.039 &  0.047 & 0.0055 & -0.31 & 0.28   \\
             & 6.31   &   6.18      & 6.29       & 6.26  & 5.92 & 7.37   &  6.77  & 6.21   & 6.27  & 6.19   \\
             & 2.40   &   2.35      & 2.37       & 2.38  & 7.74 & 2.82   &  2.54  & 3.12   & 2.39  & 2.37   \\
\hline
Set 4        & 0.35   &   -0.28     & 0.06       & 0.072 & 2.25 & -0.048 &  0.059 & 0.0075 & -0.38 & 0.34   \\
             & 5.77   &    5.62     & 5.73       & 5.77  & 6.67 & 6.85   &  7.11  & 5.80   & 5.79  & 5.70   \\
             & 1.81   &    1.74     & 1.76       & 1.82  & 16.6 & 2.09   &  2.78  & 2.60   & 1.85  & 1.82   \\
\end{tabular}
\end{table}

\begin{table}[3]
\caption{\label{table:bvg}
The value $g_+(0)$ for the transition $B\to V\gamma$.}
\centering
\begin{tabular}{|c|c|c|c|c|c|}
\multicolumn{2}{|c|}{Ref.}  & $B\to K^{*}$ & $B_{u}\to \rho$ & $B_{s}\to \phi$ & $B_{s}\to K^{*}$ \\
\hline
 &\cite{sr1}    &  0.31$\pm$ 0.04 & 0.27$\pm$ 0.04 &   &      \\
SR &\cite{sr2}    &  0.32$\pm$ 0.05 & 0.24$\pm$ 0.04 & 0.29$\pm$ 0.05 & 0.20$\pm$ 0.04   \\
 &\cite{sr3}    &  0.38$\pm$0.06 &   &   &    \\
\hline
 &\cite{lat1}  & 0.23$\pm$ 0.04  &   &   &     \\
Lat &\cite{lat2}a   & 0.32$\pm$ 0.20  &   &   &     \\
 &\cite{lat2}b   & 0.25$\pm$ 0.08  &   &   &     \\
\hline
  &\cite{jw}       & 0.31         &   &   &     \\
QM &\cite{faustov}  & 0.32$\pm$ 0.03 & 0.26$\pm$ 0.03 & 0.27$\pm$ 0.03 & 0.23$\pm$ 0.02   \\
 &\cite{stech}    & 0.35           & 0.3            &                &                  \\
 &This work       & 0.33$\pm$ 0.05 & 0.25$\pm$ 0.05 & 0.25$\pm$ 0.01 & 0.19$\pm$ 0.03   \\
\end{tabular}
\end{table}

\begin{table}[4]
\caption{\label{table:ffs0}
The form factors for the $B\to K,K^*$ transition at $q^2=0$:  $F_1=f_+$, $F_T=-(M_K+M_B)s$, 
$V=(M_{K^*}+M_B)g$, $A_1=f/(M_{K^*}+M_B)$, $A_2=-(M_{K^*}+M_B)a_+$.}
\centering
\begin{tabular}{|c|c|c|c|c|c|}
Ref.   & $F_1(0)$  & $F_T(0)$&  $V(0)$ & $A_1(0)$ & $A_2(0)$ \\ 
\hline
SR \cite{sr3} & 0.25$\pm$0.03 & -0.14$\pm$0.03 & 0.47$\pm$0.03 & 0.37$\pm$0.03 & 0.40$\pm$0.03 \\
This work & $0.33\pm0.03$ & $-0.34$  & $0.37\pm0.07$ & $0.31\pm0.05$ & $0.26\pm0.04$  
\end{tabular}
\end{table}
\end{document}